\documentclass[twocolumn,prb,showpacs]{revtex4}

\usepackage{graphicx}
\usepackage{dcolumn}
\usepackage{bm}
\usepackage{amssymb}
\usepackage{amsmath}

\begin{document}

\title{Structural characteristics of positionally-disordered lattices: 
relation to the first sharp diffraction peak in glasses}

\author{J. K. Christie}
\email{jkc25@cam.ac.uk}
\author{S. N. Taraskin}
\author{S. R. Elliott}
\affiliation{%
Department of Chemistry, Cambridge University, Lensfield Road, Cambridge, 
CB2 1EW, UK}%

\date{\today}

\begin{abstract}
Positional disorder has been introduced into the atomic structure of 
certain crystalline lattices, and the orientationally-averaged structure 
factor $S(k)$ and pair-correlation function $g(r)$ of these disordered 
lattices have been studied. Analytical expressions for $S(k)$ and $g(r)$ 
for Gaussian positional disorder in 2D and 3D are confirmed with precise 
numerical simulations. These analytic results also have a bearing on the 
unsolved Gauss circle problem in mathematics. As the positional disorder 
increases, high-\emph{k} peaks in $S(k)$ are destroyed first, eventually 
leaving a single peak, that with the lowest-\emph{k} value. The 
pair-correlation function for lattices with such high levels of positional 
disorder exhibits damped oscillations, with a period equal to the separation 
between the furthest-separated (lowest-\emph{k}) lattice planes. The last 
surviving peak in $S(k)$ is, for example for silicon and silica, at a 
wavevector nearly identical to that of the experimentally-observed first 
sharp diffraction peak (FSDP) in the amorphous phases of those materials. 
Thus, for these amorphous materials at least, the FSDP can be regarded as 
arising from scattering from atomic configurations equivalent to the single 
family of positionally-disordered local Bragg planes having the furthest 
separation.
\end{abstract}


\maketitle


Medium-range order (MRO) in the structure of amorphous and liquid
systems continues to attract a lot of attention. \cite{Elliott Nature,
Voyles and Abelson} The nature of the short-range order (SRO) in such systems 
is often accessible to diffraction or spectroscopic experiments, the physical
interpretation of which, in the form of atom-atom pair-distribution functions,
is clear. \cite{Elliott Book} The more subtle features which contribute 
to MRO, by contrast, are harder both to measure and to understand. Of 
particular interest is the so-called first sharp diffraction peak (FSDP), a 
characteristic feature of very many amorphous network materials. 
\cite{Elliott PRL} This peak occurs in the structure factor at low values 
of scattering-wavevector transfer, \emph{k} (= $\left|\mathbf{k}\right|$). 
The real-space information contained in the structure factor can be obtained 
via a Fourier transform, and so the FSDP at low \emph{k} corresponds to 
atom-atom correlations at large distances; hence the presence of the FSDP is
usually thought of as a signature of MRO. \cite{Elliott Nature} Much
effort has been expended in trying to deduce the real-space structural
features which contribute to MRO and the FSDP in various amorphous systems; 
\cite{Elliott PRL,Salmon,Nakamura et al,Wilson and Madden, Sadigh Dzugutov 
and Elliott,Massobrio and Pasquarello,Gaskell and Wallis}  
explanations have been proposed, for example,
based on the packing of voids in the material, 
\cite{Elliott PRL, Wilson and Madden}, the existence of quasi-Bragg planes,
\cite{Gaskell and Wallis} or other unknown effects.
\cite {Massobrio and Pasquarello}  
Gaskell and Wallis \cite{Gaskell and Wallis} have stated that in many 
amorphous materials (although not all), the wavevector of the FSDP is close to
the wavevector of a diffraction peak (the lowest) in a related crystalline
phase, and thus, in real space, to a set of Bragg planes (with the
largest interplanar spacing). They proposed that remnants of these
lattice planes exist in the disordered phase, and that scattering
from these planes causes the FSDP.

In this paper, we show that the wavevector of the FSDP in certain amorphous 
materials can be predicted from the lattice structure of their crystalline 
counterparts. Amorphous materials are positionally and topologically 
disordered; that is, the atoms have shifted from their crystalline 
positions. However, if the wavevector of the amorphous FSDP is close to that 
of the lowest peak in the crystalline structure factor
\cite{Gaskell and Wallis} it is reasonable to assume that the MRO 
in the amorphous phase could contain structural configurations closely 
related to the crystal.  The structure of amorphous materials was mimicked 
in this work by introducing positional disorder into the
structure of certain lattices, and the behaviour of the 
structure factor and atomic-density pair-correlation function was investigated
analytically and numerically.

Positional disorder was introduced into the crystal lattices by shifting each
atom independently a random distance away from its crystalline position.
In each of the \emph{x}-, \emph{y}- and \emph{z}-directions, the shifts were 
randomly chosen from a Gaussian distribution centred on zero, independently of
the shifts in the other directions.  The half width $\sigma$ of the Gaussian 
was thus a measure of the amount of positional disorder introduced.  
No restriction was placed on the closeness of approach of atoms (the results 
were not qualitatively different with a suitable cut-off). 

If $\sigma$ is small, the resulting disordered structure
is rather like the original crystal; if $\sigma$ is large, the 
atomic positions are located effectively randomly and little memory
remains of the underlying lattice. Figure \ref{figure, disordered diamond 
lattices} shows the structure of the diamond lattice (Fig.\ \ref{figure, 
disordered diamond lattices}a) as an
increasing amount of positional disorder is introduced. The 
underlying lattice is still very
clear at $\sigma=0.1r_{1}$ (Fig.\ \ref{figure, disordered diamond lattices}b), 
where $r_{1}$ is the crystalline nearest-neighbour distance. 
For $\sigma=0.2r_{1}$ (Fig.\ \ref{figure, disordered diamond lattices}c), some 
remnant of the \{111\} layers is still visible. Only faint vestiges of 
the crystal structure persist at $\sigma=0.3r_{1}$ (Fig.\ \ref{figure, 
disordered diamond lattices}d).

\begin{figure}[tb]
\begin{tabular}{|c|c|}
\hline 
(a)\includegraphics[%
  width=1.3in]{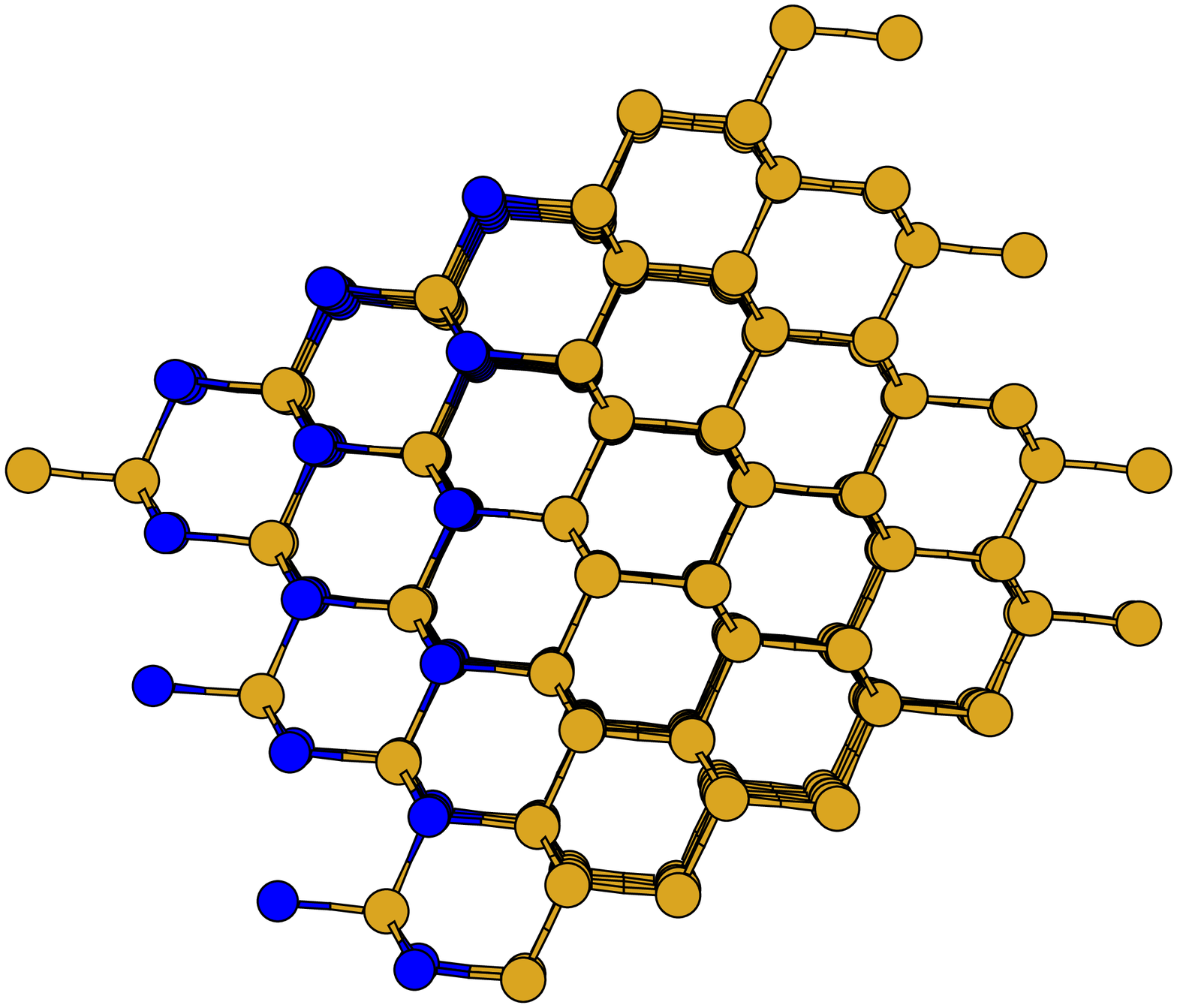}&
(b)\includegraphics[%
  width=1.3in]{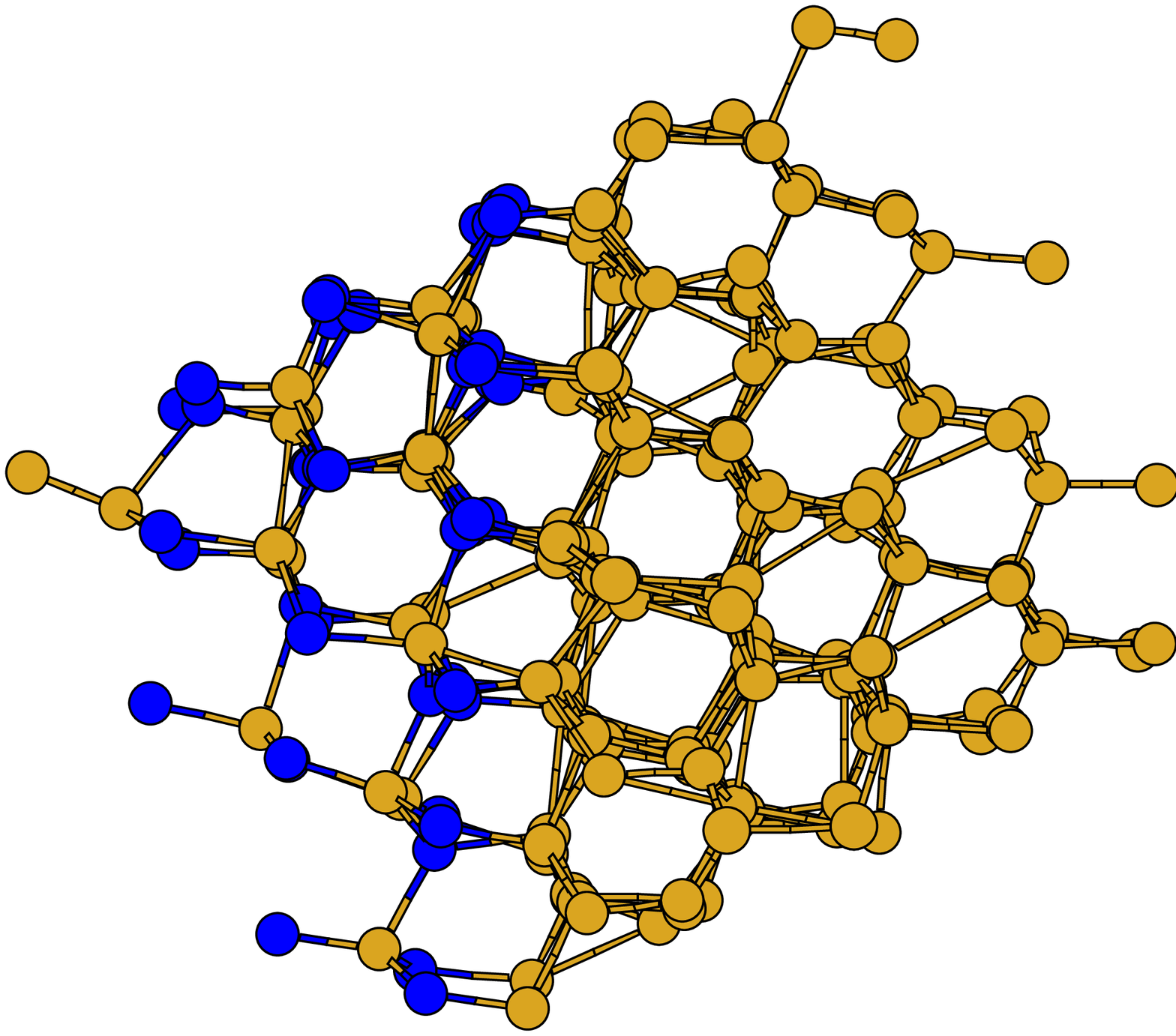}\tabularnewline
\hline
(c)\includegraphics[%
  width=1.3in]{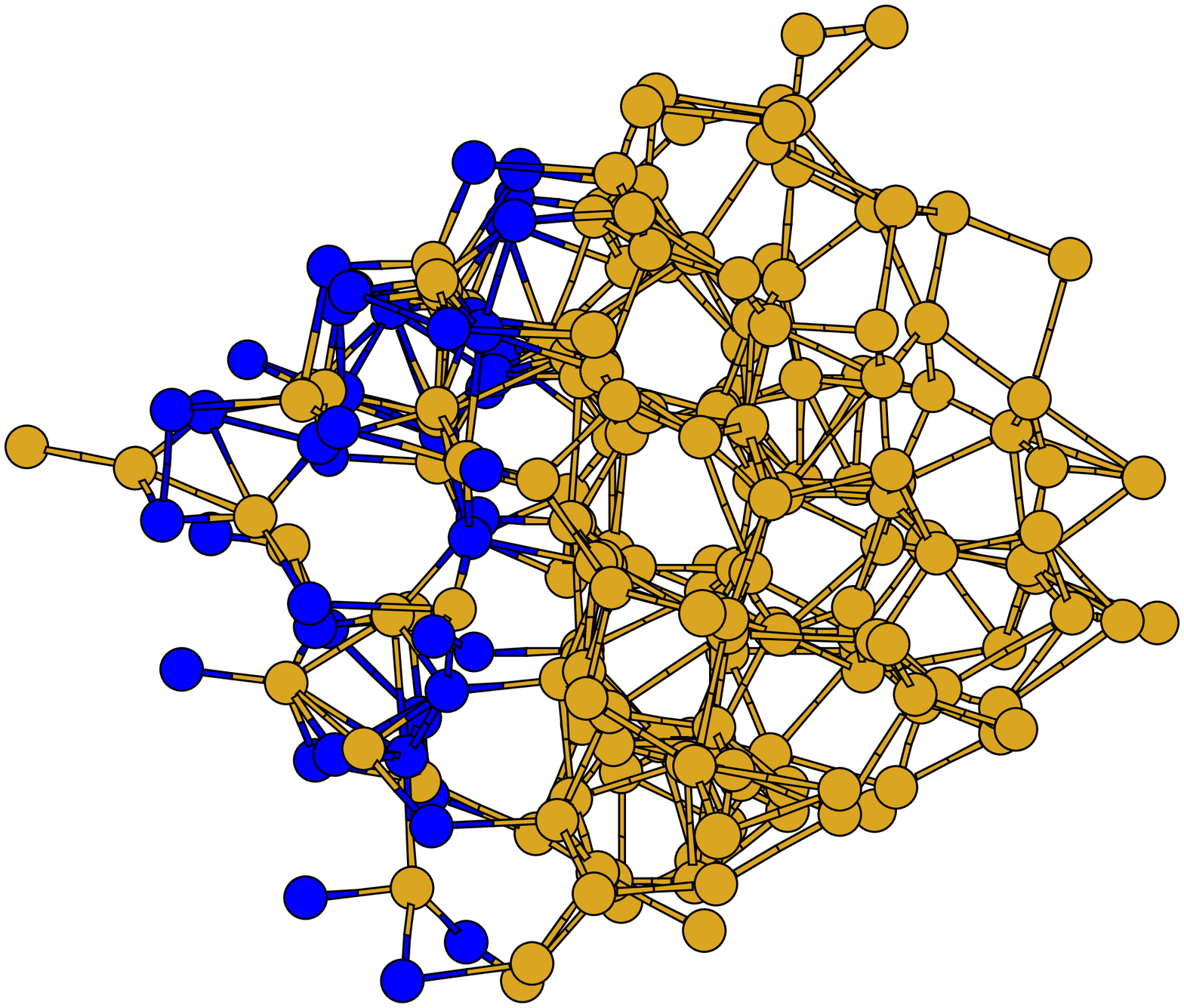}&
(d)\includegraphics[%
  width=1.3in]{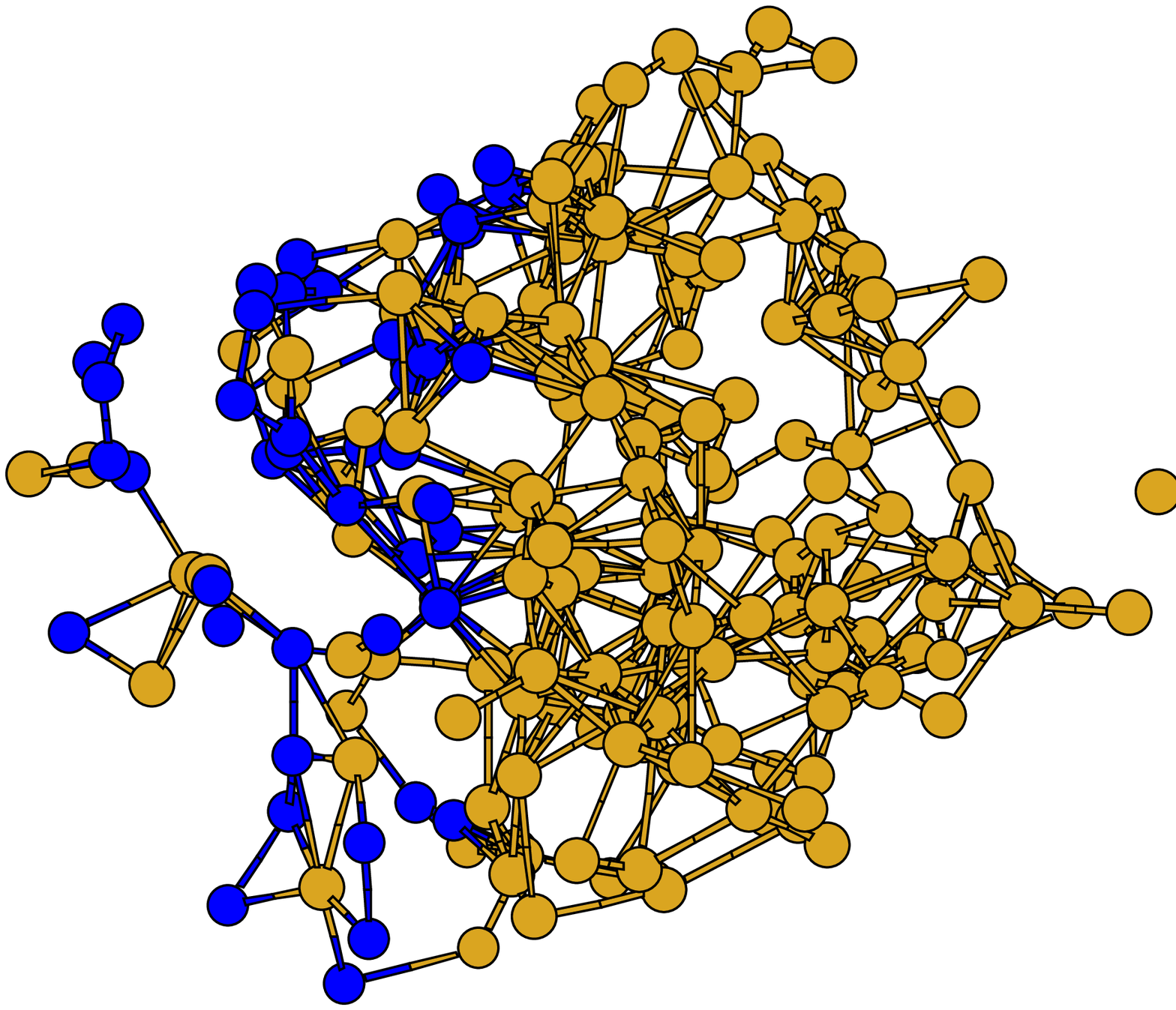}\tabularnewline
\hline
\end{tabular}
\caption{\label{figure, disordered diamond lattices}(Colour online) 
Gaussian positional disorder in the diamond lattice with values of disorder 
parameter $\sigma$: (a) 0, (b) 0.1$r_{1}$, (c) 0.2$r_{1}$, (d) 0.3$r_{1}$,
where $r_{1}$ is the nearest-neighbour distance. 
All models have the same orientation. Atoms in three consecutive \{111\} 
lattice planes perpendicular to the page are darkened to show the gradual 
destruction of the lattice planes with increasing positional disorder.}
\end{figure}

Gaussian positional disorder, as studied here, has a physical analogue: it
corresponds to thermal motion of atoms about their crystalline positions.
\cite{Egami} In this work, the displacement of the atoms from their
crystalline positions can far exceed that which would occur thermally in a
real crystal, but the mathematical formalism is identical.
In particular, the effect of thermal motion on
the structure factor $S(\mathbf{k})=1+\frac{1}{N}\sum_{i}\sum_{j\neq{i}}
\exp(\textrm{i}\mathbf{k}.\mathbf{r}_{ij})$, where $N$ is the number of 
atoms, 
and $\mathbf{r}_{ij}=\mathbf{r}_{j}-\mathbf{r}_{i}$ is the position vector
for the separation between atoms $i$ and $j$, is well-known. \cite{Willis and
Pryor, James} 
The crystalline Bragg peaks have a weight proportional to $N$;
\cite{Ashcroft and Mermin} thermal disorder reduces their weights
by the Debye-Waller factor,
$e^{-k^{2}\sigma^{2}}$ (without broadening the peaks) and a low-amplitude
diffuse scattering intensity occurs. In terms of the 
(orientationally averaged) crystalline structure factor 
$S(k)^{cryst}$, the disordered structure factor for Gaussian positional
disorder is given by:\cite{Willis and Pryor, James} 
$S(k)=1+e^{-k^{2}\sigma^{2}}(S(k)^{cryst}-1)$.
The Debye-Waller factor destroys the high-\emph{k} peaks
in the crystalline structure factor fastest as positional disorder increases
(Fig.\ \ref{structure factor, diamond lattice}).
It is clear that, as $\sigma$ increases, there will be a range of disorder
in which only one peak makes a substantial contribution to $S(k)$, the rest
being negligible; this peak will be the crystalline peak at the 
lowest value of $k$.  It is known\cite{Gaskell and Wallis} that
the FSDP often has a very similar wavevector to that of the lowest-$k$ peak 
in $S(k)^{cryst}$; here we
see that this peak is the most resistant to positional disorder. The 
lowest-$k$ peak in $S(k)$ corresponds to the set of Bragg
planes which are furthest separated in real space; they are also the set of 
planes with the highest atomic density. Figure \ref{structure factor, 
diamond lattice} shows the calculated $S(k)$ for positionally-disordered 
diamond lattices, with $N=97336$ atoms, for various values of $\sigma$. 
The peaks
decrease in height in very close accord with the theoretical Debye-Waller 
decrease.  For $\sigma\gtrsim0.3r_{1}$, only the \{111\} peak at 
$kr_{1}=4.71$ makes a substantial contribution.  

\begin{figure}
\includegraphics[%
  width=2.85in,
  angle=270]{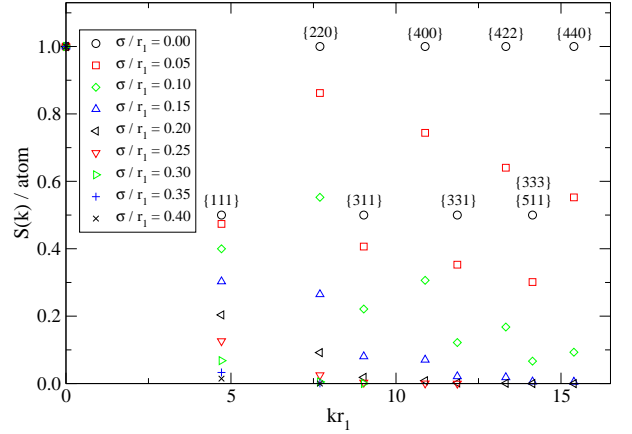}

\caption{\label{structure factor, diamond lattice}(Colour online) 
Orientationally-averaged structure factor per atom, for 
Gaussian-positionally-disordered diamond lattices (with 97336 atoms) 
with different values of disorder parameter, $\sigma$, where $r_{1}$ is the 
nearest-neighbour crystalline distance. Only the lowest-\emph{k} peak, 
at $kr_{1}=4.71$, contributes substantially for 
$\sigma\gtrsim0.3$$r_{1}$.}
\end{figure}

The structure factor $S(k)$ is related by a Fourier transform to the 
pair-correlation function $g(r)$
(the orientational average of $g(\mathbf{r})=(N\rho_{0})^{-1}\sum_{i}
\sum_{j\neq i}\delta\left(\mathbf{r}-\mathbf{r}_{ij}\right)$, where
$\rho_{0}$ is the average density). Thus,
the effect of positional disorder is to convolve the $\delta$-function 
peaks in the crystalline $g(\mathbf{r})$ with the Fourier transform of the 
Debye-Waller factor.
For Gaussian disorder, this gives:\begin{equation}
g(\mathbf{r})=\frac{1}{(2\sqrt{\pi})^{d}\sigma^{d}N\rho_{0}}
\sum_{i}\sum_{j\neq{i}}\exp\{\frac{-(\mathbf{r}-\mathbf{r}_{ij})^{2}}
{4\sigma^{2}}\},\label{eq:g(r)}\end{equation}
where $d$ is the dimensionality of the system.

In principle, $g(r)$ can be evaluated by summing eq.\ \ref{eq:g(r)}
over all crystalline
interatomic distances for a given lattice, and orientationally averaging the
result.  For the 2D square lattice with crystalline lattice positions 
$\mathbf{r}_{i}=n_{x}\mathbf{\hat x}+n_{y}\mathbf{\hat y}$, $g(\mathbf{r})$ 
in eq.\ \ref{eq:g(r)}
can be orientationally averaged to give:
\begin{eqnarray}
g(r)=\frac{1}{4\pi\sigma^{2}\rho_{0}}\sum_{n_{x},n_{y}=-\infty}^{\infty}
\exp(-(r^{2}+n_{x}^{2}+n_{y}^{2})/4\sigma^{2})\times\nonumber\\
I_{0}\left(\frac{r\sqrt{n_{x}^{2}+n_{y}^{2}}}{2\sigma^{2}}\right),
\label{eq:gr2D}
\end{eqnarray}
where $I_{0}(z)$ is a modified Bessel function of the first kind. For the 
3D simple cubic lattice, with crystalline lattice positions 
$\mathbf{r}_{i}=n_{x}\mathbf{\hat x}+n_{y}\mathbf{\hat y}
+n_{z}\mathbf{\hat z}$,
$g(r)$ is given by:
\begin{eqnarray}
g(r)=\frac{1}{8\pi^{3/2}\sigma\rho_{0}r}\times \nonumber\\
\mathop{\sum_{n_{x},n_{y},n_{z}=-\infty}^{\infty}}
\frac{e^{-(r-n)^{2}/4\sigma^{2}}-e^{-(r+n)^{2}/4\sigma^{2}}}
{n},
\label{eq:gr3D}
\end{eqnarray}
where $n=\sqrt{n_{x}^{2}+n_{y}^{2}+n_{z}^{2}}$, and the sums in 
eqs.\ \ref{eq:gr2D},\ref{eq:gr3D} exclude the
case where all indices $n_{i}$ are zero simultaneously.  It has not been 
possible to evaluate these sums analytically. In retrospect, this failure 
is unsurprising.  More than 150 years ago, Gauss first posed what is
known as the Gauss circle problem:\cite{Guy} given a regular square lattice 
of points, and a circle of radius $r$ centred on one 
of the points, how many lattice points lie within the circle? The exact answer 
remains unknown, despite much attention. \cite{Huxley} An exact 
expression for $g(r)$ with a width $\sigma$ of zero would solve this problem.
The equivalent problem is also unsolved in more than two dimensions.

Since it has not been possible to obtain $g(r)$ analytically, it was 
calculated numerically by simulating large ($\sim10^{5}$ atoms) crystalline 
models of 2D square and 3D simple cubic (SC), face-centred cubic (FCC), 
body-centred cubic (BCC), diamond, 
and cristobalite (SiO$_{\textrm{2}}$) structures, which were then 
Gaussian positionally disordered with various values of $\sigma$.
Periodic boundary conditions (PBCs) were used, and it was necessary to use 
large models in order to increase the number of allowed \textbf{k}-vectors, 
particularly for low \emph{k}. PBCs also imply that $g(r)$ is not
reliable for distances beyond half of the (cubic) box size $L$; data
are not shown for values of $r>L/2$.

For the crystal (with $\sigma=0$), $g(r)$ consists of a series of 
$\delta$-functions at the various interatomic distances. For small $\sigma$
(for $\sigma\lesssim0.2r_{1}$ - not 
shown here), the $\delta$-functions broaden into Gaussians \cite{Egami} and 
merge, giving a complicated form. However, as $\sigma$ is 
further increased, $g(r)$ takes on a simple oscillatory form with a 
\emph{single} frequency, having an envelope that decays spatially
as a power law (see Fig.\ \ref{atomic density, diamond, single frequency}).
The 3D lattices were all found to have $g(r)$ decaying as $r^{-1}$, whilst 
the 2D square lattice was found to have $g(r)$ decaying as 
$r^{-1/2}$. The reason for this is clear: the 3D Gaussian $g(r)$
(eq.\ \ref{eq:gr3D}) has an explicit $r^{-1}$ dependence, and the 
asymptotic behaviour of the modified Bessel function in the 2D $g(r)$ 
(eq.\ \ref{eq:gr2D}) has an $r^{-1/2}$ dependence. The
simulated $g(r)$'s (Fig.\ \ref{atomic density, diamond, 
single frequency}) agree very closely with the theoretical expressions given
by eqs.\ \ref{eq:gr2D},\ref{eq:gr3D}. If $g(r)$ consists of oscillations 
with a single significant frequency of period $D$, then there is only 
one significant peak in the structure factor $S(k)$, at a wavevector of 
$k_{1}=2\pi/D$. A heuristic argument for why only a single period is
observed in the oscillatory behaviour of $g(r)$ at sufficiently large $\sigma$
draws on the Rayleigh criterion for resolving overlapping broad peaks, such
as those that occur in the sums of Gaussians in 
eqs.\ \ref{eq:gr2D},\ref{eq:gr3D}: two peaks are just resolvable when
the full width of the peaks is comparable to their separation. Thus, for
sufficiently large $\sigma$, only contributions from lattice planes with the
\emph{largest} separation combine to give resolvable oscillations in $g(r)$.

The periods of the oscillations, $D$, were extracted by assuming the 
empirical form $g(r)=1+Ar^{-1}\sin(k_{1}r+\phi)$ for the 3D 
lattices, and $g(r)=1+Ar^{-1/2}\sin(k_{1}r+\phi)$ for the square
lattice, and performing a least-squares minimisation with respect
to the variables \emph{A}, $k_{1}$ and $\phi$. Non-oscillatory data 
for $r<2r_{1}$ were excluded from the
fit. Averages were then taken over twenty realisations of each
disordered model. For all the lattices studied, $D$ is equal to the spacing 
between the furthest-separated 
lattice planes (see Table \ref{table, planes}).  We conclude that these 
single-frequency oscillations in $g(r)$ correspond to the Fourier transform 
of the single remaining peak in $S(k)$. 

\begin{table*}
\begin{tabular}{|c|c|c|c|}
\hline 
Lattice&
Furthest planes&
Interplanar distance&
Period of oscillations, \emph{D}\tabularnewline
\hline
SC&
{[}100{]}&
$r_{1}$&
$\left(1.0000\pm0.0003\right)r_{1}$ for $\sigma=0.25r_{1}$\tabularnewline
\hline
FCC&
{[}111{]}&
$\sqrt{\frac{2}{3}}$$r_{1}=0.8165r_{1}$ &
$\left(0.8167\pm0.0001\right)r_{1}$ for $\sigma=0.25r_{1}$\tabularnewline
\hline 
BCC&
{[}110{]}&
$\sqrt{\frac{2}{3}}$$r_{1}=0.8165r_{1}$ &
$\left(0.8165\pm0.0001\right)r_{1}$ for $\sigma=0.25r_{1}$\tabularnewline
\hline 
Diamond&
{[}111{]}&
$\frac{4}{3}r_{1}=1.3333r_{1}$&
$\left(1.3333\pm0.0001\right)r_{1}$ for $\sigma=0.35r_{1}$\tabularnewline
\hline 
Square&
{[}10{]}&
$r_{1}$ &
$\left(0.99993\pm0.00001\right)r_{1}$ for $\sigma=0.3r_{1}$\tabularnewline
\hline
\end{tabular}

\caption{\label{table, planes}Families of lattice planes with the largest
interplanar distance for certain crystal structures; $r_{1}$ is the
nearest-neighbour distance in each case. For the values of 
disorder parameter, $\sigma$, given,
$g(r)$ consists of oscillations having only a single frequency, the periods of
which, \emph{D}, were found from a best-fit procedure, as described in 
the text.}
\end{table*}

\begin{figure}
\includegraphics[%
  width=2.75in,
  angle=270]{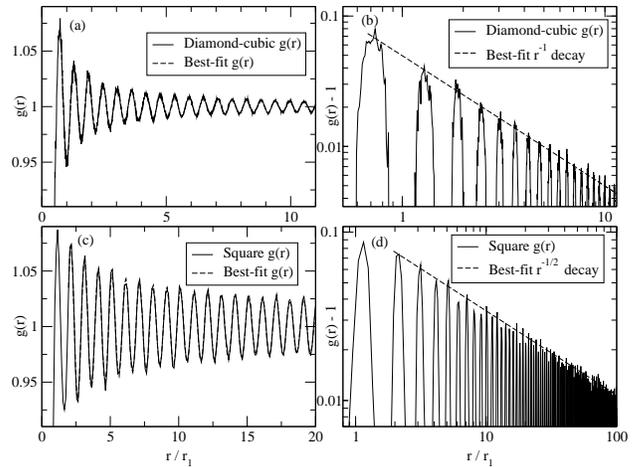}

\caption{\label{atomic density, diamond, single frequency}(a) $g(r)$ for a 
Gaussian-positionally-disordered diamond lattice with disorder
parameter $\sigma=0.3r_{1}$ (solid), and the best-fit curve (dashed). 
(b) The same data plotted double-logarithmically with the 
best-fit $r^{-1}$ decay envelope. (c) $g(r)$ for a 
Gaussian-positionally-disordered square lattice with
$\sigma=0.25r_{1}$ (solid), and the best-fit curve
(dashed). (d) The same data plotted double-logarithmically with the 
best-fit $r^{-1/2}$ decay envelope. Both best-fit curves exclude the
non-oscillatory data at low $r$.}
\end{figure}

We now show that positionally-disordered
structures of appropriate crystal lattices can predict the position
of the FSDP in various amorphous systems. Real amorphous materials show several
peaks in $S(k)$, all of which comprise contributions from many
interatomic distances. The corresponding $g(r)$ has several aperiodic peaks,
and the single-period oscillation shown in Fig.\ \ref{atomic density, 
diamond, single frequency} would modulate such peaks, giving rise to the FSDP.
Recently, {}``time-frequency'' wavelet-transform
analysis has been applied to the study of disordered structures to
elucidate the interatomic-distance contributions to different peaks,
particularly the FSDP. \cite{Harrop Taraskin Elliott,Harrop PhD,Uchino et al}
We show below that the FSDP in $S(k)$ for some amorphous
materials could be due to scattering from structural
configurations equivalent to the last surviving family of local, disordered 
Bragg planes in a positionally-disordered structure.

Silicon crystallises in the diamond structure,
and its amorphous form has been well studied as a canonical example
of a topologically-disordered solid.\cite{Elliott Book} 
It also exhibits a strong FSDP at $k_{1}=2.00$\AA$^{-1}$.\cite{Kugler et al}
From Table \ref{table, planes}, the period of the last surviving
extended-range oscillations in $g(r)$ for the positionally-disordered
diamond lattice is the distance between successive \{111\} planes,
i.e. the apex--basal-plane distance in an SiSi$_{\textrm{4}}$ tetrahedron,
$D_{111}=\frac{4}{3}r_{1}$. In crystalline silicon, 
$r_{1}=2.35$\AA, \cite{CRC} corresponding to an oscillation
with a period of $D=3.13$\AA, and hence $k_{1}=2\pi/3.13$\AA$=2.01$\AA$^{-1}$,
in excellent agreement with the experimental value above. \cite{Kugler et al}

The FSDP is often observed in AX$_{\textrm{2}}$-type glasses, 
\cite{Elliott Nature,Elliott PRL}
the most well-known example being vitreous silica, v-SiO$_{\textrm{2}}$.
A crystalline counterpart of v-SiO$_{\textrm{2}}$ is the tetragonal
$\alpha$-cristobalite structure. \cite{Tucker et al}
Real v-SiO$_{\textrm{2}}$ is known \cite{Elliott Nature}
to consist of fairly ordered SiO$_{\textrm{4}}$ tetrahedra, linked
in a corner-sharing manner in a disordered network; we made no attempt
to keep the tetrahedra ordered, and Gaussian disordered the atomic positions
of cristobalite
without restriction. Hence, these models are not a particularly
good representation of real v-SiO$_{\textrm{2}}$.
Nevertheless, single-frequency extended-range oscillations in $g(r)$ 
were found in positionally-disordered cristobalite structures for 
$0.35\lesssim\sigma/r_{1}\lesssim0.6$.
The period of this oscillation is $D\simeq4.07$\AA,
corresponding to a peak in $S(k)$ at $k_{1}\simeq1.54$\AA$^{-1}$,
in very good agreement with the experimental value of the
FSDP wavevector at $k_{1}=1.52$\AA$^{-1}$. \cite{Susman et al PRB}

In conclusion, Gaussian positional disorder has been introduced into a variety
of crystalline lattices, specifically, the 3D diamond, FCC, BCC, SC and the 
2D square
lattices, as well as the cristobalite crystalline polymorph of silica.
As this positional disorder increases, the high-\emph{k} Bragg peaks
in the structure factor are destroyed first, eventually leaving only
the lowest-\emph{k} peak in $S(k)$. Examination of the 
orientationally-averaged real-space 
pair-correlation function of such disordered structures shows power-law-damped,
single-frequency, extended-range oscillations, with a period, \emph{D},
equal to the spacing between the most distant (most areally dense)
lattice planes. This period is equal to $D=2\pi/k_{1}$, where $k_{1}$
is the wavevector of the first peak in the structure factor. The atomic 
positions of crystalline counterparts of silicon
and silica have been disordered. 
The wavevector of the FSDP in real amorphous silicon
and in vitreous silica, is nearly identical to that of the lowest-$k$ 
peak in the structure factor of the
positionally-disordered crystalline counterparts,
and hence at sufficiently large distances from an arbitrary origin atom,
the atomic structure of a topologically-disordered amorphous material appears
to be similar to that of the corresponding positionally-disordered crystal.

JKC would like to thank the Engineering and Physical Sciences Research
Council for financial support.


\begin{thebibliography}{10}
\bibitem{Elliott Nature}S.\ R.\ Elliott, Nature \textbf{354}, 445 (1991)
\bibitem{Voyles and Abelson}P.\ M.\ Voyles and J.\ R.\ Abelson, Solar Energy 
Materials \& Solar Cells \textbf{78}, 85 (2003)
\bibitem{Elliott Book}S.\ R.\ Elliott, \emph{Physics of Amorphous Materials}, 
(Longman, 1990)
\bibitem{Elliott PRL}S.\ R.\ Elliott, Phys.\ Rev.\ Lett.\ \textbf{67}, 
711 (1991)
\bibitem{Salmon}P.\ S.\ Salmon, Proc.\ R.\ Soc.\ Lond.\ A \textbf{445}, 3
51 (1994)
\bibitem{Nakamura et al}M.\ Nakamura, M.\ Arai, Y.\ Inamura, T.\ Otomo and 
S.\ M.\ Bennington, Phys.\ Rev.\ B \textbf{67}, 064204 (2003)
\bibitem{Wilson and Madden}M.\ Wilson and P.\ A.\ Madden, Phys.\ Rev.\ Lett.\ 
\textbf{80}, 532 (1998)
\bibitem{Sadigh Dzugutov and Elliott}B.\ Sadigh, M.\ Dzugutov and 
S.\ R.\ Elliott, Phys.\ Rev.\ B \textbf{59}, 1 (1999)
\bibitem{Massobrio and Pasquarello}C.\ Massobrio and A.\ Pasquarello, 
J.\ Chem.\ Phys.\ \textbf{114}, 7976 (2001)
\bibitem{Gaskell and Wallis}P.\ H.\ Gaskell and D.\ J.\ Wallis, 
Phys.\ Rev.\ Lett.\ \textbf{76}, 66 (1996)
\bibitem{Egami}T.\ Egami, Z.\ Kristallogr.\ \textbf{219}, 122 (2004)
\bibitem{Willis and Pryor}B.\ T.\ M.\ Willis and A.\ W.\ Pryor, \emph{Thermal 
Vibrations in Crystallography}, (Cambridge University Press, 1975)
\bibitem{James}R.\ W.\ James, \emph{The Optical Principles of the Diffraction 
of X-rays}, (G.\ Bell \& Sons, 1965)
\bibitem{Ashcroft and Mermin} N.\ W.\ Ashcroft and N.\ D.\ Mermin,
\emph{Solid State Physics}, (International Thomson Publishing, 1976),
Appendix F
\bibitem{Guy}R.\ K.\ Guy, \emph{Unsolved Problems in Number Theory}, 
(Springer-Verlag, 1994), Sec.\ F1
\bibitem{Huxley}M.\ N.\ Huxley, \emph{Area, Lattice Points and Exponential 
Sums}, (Oxford University Press, 1996)
\bibitem{Harrop Taraskin Elliott}J.\ D.\ Harrop, S.\ N.\ Taraskin and 
S.\ R.\ Elliott, Phys.\ Rev.\ E \textbf{66}, 026703 (2002); \textbf{68}, 
019904(E) (2003)
\bibitem{Harrop PhD}J.\ D.\ Harrop, Ph.D.\ thesis, Cambridge University, 2004
\bibitem{Uchino et al}T.\ Uchino, J.\ D.\ Harrop, S.\ N.\ Taraskin and 
S.\ R.\ Elliott, (unpublished)
\bibitem{Kugler et al}S.\ Kugler, L.\ Pusztai, L.\ Rosta, P.\ Chieux and 
R.\ Bellissent, Phys.\ Rev.\ B \textbf{48}, 7685 (1993)
\bibitem{CRC}\emph{CRC Handbook of Chemistry and Physics}, edited by 
D.\ R.\ Lide, (CRC Press, 2003), 84th edn.
\bibitem{Tucker et al}M.\ G.\ Tucker, M.\ P.\ Squires, M.\ T.\ Dove and 
D.\ A.\ Keen, J.Phys.: Condens.\ Matter \textbf{13}, 403 (2001)
\bibitem{Susman et al PRB}S.\ Susman, K.\ J.\ Volin, D.\ L.\ Price, 
M.\ Grimsditch, J.\ P.\ Rino, R.\ K.\ Kalia, P.\ Vashishta, G.\ Gwanmesia, 
Y.\ Wang and R.\ C.\ Liebermann, Phys.\ Rev.\ B \textbf{43}, 1194 (1991)
\end{thebibliography}
\end{document}